\documentclass[twocolumn]{aastex62}

\usepackage{CJK}
\usepackage{comment}
\usepackage{hyperref}

\usepackage{txfonts}
\usepackage{color,bm}
\usepackage [latin1]{inputenc}

\newcommand{\eqref}[1]{(\ref{#1})}

\shorttitle{Dusty Rings as a Cause of Impossible Spectral Slopes}
\shortauthors{Ohno, Thao, Mann, Fortney}

\begin{document}

\title{A Circumplanetary Dust Ring May Explain the Extreme Spectral Slope of the $10~{\rm Myr}$ Young Exoplanet K2-33b}

\author[0000-0003-3290-6758]{Kazumasa Ohno}
\affil{Department of Astronomy \& Astrophysics, University of California, Santa Cruz, 1156 High St, Santa Cruz, CA 95064, USA}

\author[0000-0001-5729-6576]{Pa Chia Thao}
\altaffiliation{NSF Graduate Research Fellow}
\altaffiliation{Jack Kent Cooke Foundation Graduate Scholar}
\affiliation{Department of Physics and Astronomy, The University of North Carolina at Chapel Hill, Chapel Hill, NC 27599, USA} 

\author[0000-0003-3654-1602]{Andrew W. Mann}
\affiliation{Department of Physics and Astronomy, The University of North Carolina at Chapel Hill, Chapel Hill, NC 27599, USA}

\author[0000-0002-9843-4354]{Jonathan J. Fortney}
\affil{Department of Astronomy \& Astrophysics, University of California, Santa Cruz, 1156 High St, Santa Cruz, CA 95064, USA}

\begin{abstract}
Young exoplanets are attractive targets for atmospheric characterization to explore the early phase of planetary evolution and the surrounding environment. 
Recent observations of the 10 Myr young Neptune-sized exoplanet K2-33b revealed that the planet's transit depth drastically decreases from the optical to near-infrared wavelengths. Thao et al. (2022) suggested that a thick planetary haze and/or stellar spots may be the cause; however, even the best-fit model only barely explains the data. 
Here, we propose that the peculiar transmission spectrum may indicate that K2-33b possess a circumplanetary dust ring, an analog of Jupiter's dust ring. 
We demonstrate that the ring could produce a steep slope in the transmission spectrum even if its optical depth is as low as $\sim{10}^{-2}$.
We then apply a novel joint atmosphere-ring retrieval to K2-33b and find that the ring scenario could well explain the observed spectrum for various possible ring compositions.
Importantly, the dust ring also exhibits prominent absorption features of ring particles around $\sim10~{\rm {\mu}m}$, whose shape and strength depend on the composition of the ring.
Thus, future observations by JWST-MIRI would be able to test not only the ring hypothesis but also, if it indeed exists, to constrain the composition of the ring --- providing a unique opportunity to explore the origins of the dust ring around its parent planet, soon after the planetary system's formation.
\end{abstract}

\keywords{Exoplanet atmospheres (487); Exoplanet atmospheric composition (2021); Exoplanet rings (494); Transmission spectroscopy (2133)}

\section{Introduction} \label{sec:intro}
Transmission spectroscopy is a widely used method to explore the physical and chemical nature of exoplanetary atmospheres.
The spectral features of the gas molecules tell us the chemical composition of exoplanets, which provide valuable constraints on atmospheric physics, chemistry, and planet formation \citep[for recent reviews, see][]{Madhusudhan19,Zhang20,Guillot+22}.
The ``spectral slope'', the gradual increase in the transit radius toward the blue, also encapsulates information on atmospheric clouds and hazes \citep{Lecavelier+08,Pinhas&Madhusudhan17,Ohno&Kawashima20}.
Given the success in probing previously unexplored wavelength ranges by JWST \citep{ERS+22}, transmission spectroscopy will continue to be an important observational method for exoplanetary atmospheres.

Young exoplanets ($<$ 1 Gyr) are attractive targets for studying atmospheric processes and surrounding environments in the early phase of planetary evolution.
One of the best examples is K2-33b, a super-Neptune-sized (R$\sim$5R$_{\oplus}$) exoplanet orbiting a pre-main-sequence ($\sim$10 Myr old) M dwarf with an orbital period of 5.424 days \citep{Mann+16,David+16}.
Recently, Thao et al. (2022) obtained the transmission spectrum of K2-33b by combining transit observation data of {\it K2}, MEarth, the {\it Spitzer Space Telescope}, and the {\it Hubble Space Telescope}. They found that its transit depth suddenly decreases from optical to near-infrared wavelengths by a factor of $\sim2$.  Such a steep spectral slope is typically attributed to unocculted stellar spot \citep[e.g.,][]{McCullough+14,Louden+17} and/or aerosol opacity \citep[e.g.,][]{Pinhas&Madhusudhan17,Ohno&Kawashima20}. 
However, Thao et al. (2022) found that the stellar spot alone cannot explain the extreme slope of K2-33b, as the model requires a spot coverage ($71\%$) much larger than limits from the stellar spectrum ($\la$20\%).
Thao et al. (2022) argued that photochemical haze might explain the extreme slope if the planetary mass is less than $\sim 5M_{\rm \oplus}$ based on the haze formation framework of \citet{Gao&Zhang20}. 
However, \citet{Kubyshkina+18} performed hydrodynamical simulations of atmospheric escape from K2-33b and suggested that the mass should be $> 7M_{\rm \oplus}$; otherwise, the planet would have lost its atmosphere through hydrodynamic escape even at the $\sim10~{\rm Myr}$ age.

In this letter, we propose an alternative scenario that may explain the peculiar transmission spectrum of K2-33b: a circumplanetary dust ring (CPDR hereafter).
All giant planets in the Solar System are known to possess tenuous CPDR that consist of micron-sized dust particles (see \citealt[][]{Burns+01} and references therein).
A lifetime of the circumplanetary small dust is short: the orbital decay timescale of micron-sized dust is $\la0.1$ Myr for Jovian ring \citep{Burns+01} and even shorter for close-in exoplanets because of a strong Poynting-Robertson drag \citep[][]{Schlichting&Chang11}. Thus, CPDRs are believed to be sustained by continuous replenishment of dust particles, such as the capture of interplanetary dust \citep[e.g.,][]{Colwell+98,Mitchell+05} and impact ejecta from circumplanetary objects, e.g., moons \citep[e.g.,][]{Burns+99,Kruager+99,Kruger+00,Krivov+02}. 
In particular, the latter has been suggested as a viable origin of Jovian dust ring \citep[for review, see][]{Burns+01}.
Since K2-33 is an infant system ($\sim10$ Myr), it may be experiencing more CPDR-forming impact events than those in the Solar System due to the potential presence of a debris disk, and/or perhaps the remnant of a circumplanetary disk.
In Section \ref{sec:method_summary}, we use analytic and numerical models to demonstrate that a moderately optically thin CPDR could cause an extreme spectral slope.
In Section \ref{sec:spectrum_K2_33b}, we interpret the transmission spectrum of K2-33b by applying a joint atmosphere-ring retrieval that simultaneously constrains the properties of the planetary atmospheres and the CPDR in Bayesian framework. 
In Section \ref{sec:summary}, we summarize our findings.

\section{The Idea}\label{sec:method_summary}

\subsection{Spectral Slope Generated by Optically Thin Ring}
\begin{figure*}[t]
\centering
\includegraphics[clip, width=\hsize]{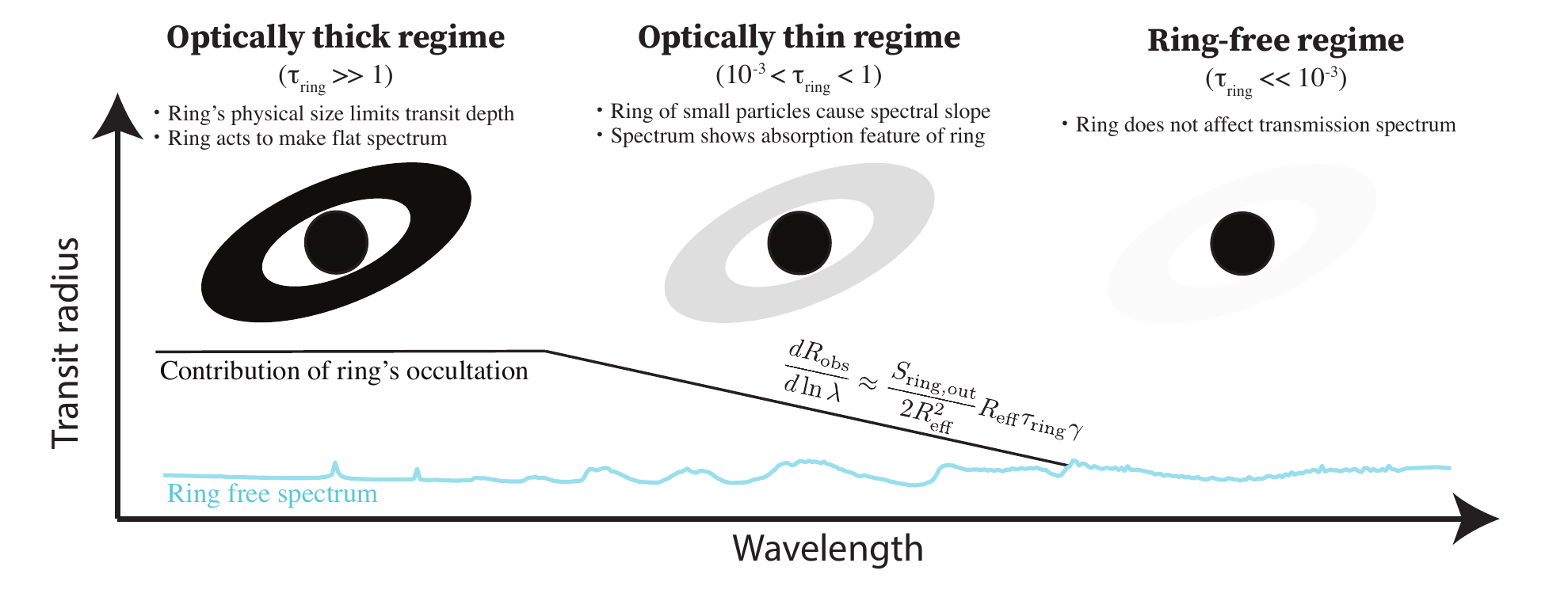}
\caption{Schematic illustration showing how the circumplanetary ring affects the atmospheric transmission spectrum. }
\label{fig:cartoon}
\end{figure*}
We first analytically demonstrate that CPDRs can potentially cause an extremely steep spectral slope in an exoplanetary transmission spectrum.
While previous studies showed that the circumplanetary ring acts to flatten the transmission spectrum \citep{Ohno&Tanaka21,Ohno&Fortney22,Alam+22}, the ring could produce a steep spectral slope if the ring is optically thin.
As shown in \citet{Ohno&Fortney22}, the transit depth of the ringed exoplanet can be well approximated by
\begin{equation}\label{eq:ring_prescription}
    \pi R_{\rm obs}^2=\pi R_{\rm eff}^2 + S_{\rm ring,out}(1-e^{-\tau_{\rm ring}})
\end{equation}
where $R_{\rm eff}$ is the effective transit radius of a planet without a ring, $S_{\rm ring,out}$ is the area of the ring projected onto the stellar disk subtracted by the area overlapped with the planetary disk, and $\tau_{\rm ring}$ is the line-of-sight optical thickness of the ring.
Differentiating this equation with respect to wavelength, one obtains the gradient of spectral slope as \citep[Eq 44 of][]{Ohno&Fortney22}
\begin{equation}\label{eq:dRdlmd}
    \frac{dR_{\rm obs}}{d\lambda}=\frac{ \frac{dR_{\rm eff}}{d\lambda} + \frac{S_{\rm ring,out}e^{-\tau_{\rm ring}} }{2R_{\rm eff}}\frac{d\tau_{\rm ring}}{d\lambda} }{\left[1+\frac{S_{\rm ring}(1-e^{-\tau_{\rm ring}})}{\pi R_{\rm eff}^2}\right]^{1/2}},
\end{equation}
where we have assumed that $dS_{\rm ring,out}/d\lambda=0$ for simplicity, which is valid if the projected minor axis of the inner edge of the ring is larger than the planetary radius.
For $\tau_{\rm ring}\ll1$, with an approximation of $e^{-\tau_{\rm ring}}\approx1$ for clarity, Equation \eqref{eq:dRdlmd} can be approximated by
\begin{equation}
    \frac{dR_{\rm obs}}{d\lambda}\approx \frac{dR_{\rm eff}}{d\lambda}+ \frac{S_{\rm ring,out}}{2R_{\rm eff}}\frac{d\tau_{\rm ring}}{d\lambda}.
\end{equation}
In this Equation, the first term expresses the atmosphere contribution, while the second term expresses the ring contribution.
Assuming a power-law opacity of the ring as $\tau_{\rm ring}\propto \lambda^{\gamma}$, we obtain
\begin{equation}\label{eq:slope_ring}
    \frac{1}{H}\frac{dR_{\rm obs}}{d\ln{\lambda}}\approx  \frac{\alpha}{1-\beta} +\frac{S_{\rm ring,out}}{2R_{\rm eff}^2}\left(\frac{R_{\rm eff}}{H}\right)\tau_{\rm ring}\gamma,
\end{equation}
where we have used $dR_{\rm eff}/d\ln{\lambda}=H\alpha/(1-\beta)$ for the atmospheric component that is derived by assuming the atmospheric opacity of $\kappa \propto \lambda^{\alpha}P^{-\beta}$ with $\beta<1$ \citep{Ohno&Kawashima20}.

Importantly, the slope generated by the ring is scaled by the projection area of the ring $S_{\rm ring,out}$ whose length scale is comparable to the planetary radius.
Thus, the ring potentially produces a spectral slope much steeper than that expected for ring-free planets, which is scaled by the atmospheric scale height.
The ring may have an outer edge around the Roche radius \citep[e.g.,][]{Piro&Vissapragada19} given by 
\begin{equation}\label{eq:roche}
    R_{\rm roche} \approx 2.46 R_{\rm p}\left( \frac{\rho_{\rm p}}{\rho_{\rm r}}\right)^{1/3}=2.46\left( \frac{3M_{\rm p}}{4\pi \rho_{\rm r}}\right)^{1/3},
\end{equation}
where $R_{\rm p}\approx R_{\rm eff}$ is the planetary radius, $\rho_{\rm p}$ is the planetary bulk density, and $\rho_{\rm r}$ is the density of the ring particles.
The effect of the ring is maximized in the extreme case of a face-on ring with a ring's inner edge close to the planetary radius, i.e., $S_{\rm ring,out}=\pi (R_{\rm roche}^2 -R_{\rm p}^2)\sim \pi R_{\rm roche}^2$.
Thus, we may crudely evaluate the maximal effect of the ring component as
\begin{eqnarray}\label{eq:ring_slope}
    \nonumber
    \frac{1}{H}\frac{dR_{\rm obs}}{d\ln{\lambda}}&\sim& \frac{ \pi R_{\rm roche}^2}{2R_{\rm eff}^2}\left(\frac{R_{\rm eff}}{H}\right)\tau_{\rm ring}\gamma\\
    &\sim& 1700 \tau_{\rm ring}\gamma \left( \frac{\rho_{\rm p}}{\rho_{\rm r}} \right)^{2/3} \left( \frac{H}{400~{\rm km}} \right)^{-1}\left( \frac{R_{\rm eff}}{R_{\rm Jupiter}} \right).
\end{eqnarray}
Equation \eqref{eq:ring_slope} indicates that the ring could cause a steep slope if the optical depth is greater than $\sim {10}^{-3}$.
For example, for $\gamma=-4$ and $\rho_{\rm p}/\rho_{\rm r}=1$, even an optically thin ring of $\tau_{\rm ring}={10}^{-2}$ yields a spectral slope of $H^{-1}dR_{\rm eff}/d\ln{\lambda}\sim -68$!
Thus, Equation \eqref{eq:ring_slope} demonstrates that the optically thin CPDR has the ability to produce an extreme spectral slope.
Of course, the above estimate is based on a highly simplified case of a face-on ring. 
The actual ring-generated slope depends on the ring's properties, such as inclination.

Figure \ref{fig:cartoon} schematically summarizes how the ring affects the transmission spectrum. If the ring is optically thick, as shown in previous studies \citep{Ohno&Fortney22,Alam+22}, the ring acts to produce a flat transmission spectrum regardless of its optical properties.
A moderately optically thin ring, say ${10}^{-3}\la \tau_{\rm ring}\la 1$, significantly affects the shape of the transmission spectrum, depending on the ring's optical properties.
If the ring's opacity has a power-law dependence on wavelength, the ring produces a spectral slope that can be much steeper than that produced by atmospheric processes (Equation \ref{eq:ring_slope}).
As shown later, an optically thin ring also produces an absorption feature of the ring particles in the spectrum (see also \citealt{Ohno&Fortney22}).
When the optical depth of a ring is too low, say $\tau_{\rm ring}\ll {10}^{-3}$, a ring does not affect the transmission spectrum.
We note that a whole spectrum potentially contains multiple regimes because the ring's optical depth can depend on wavelength. For example, if the ring optical depth suddenly decreases from 10 at visible to 0.1 at near-infrared wavelengths, a planet could show a flat spectrum at visible, a ring-free atmospheric spectrum at near-infrared, and a steep slope at intermediate wavelengths. 

\begin{figure}[t]
\centering
\includegraphics[clip, width=0.99\hsize]{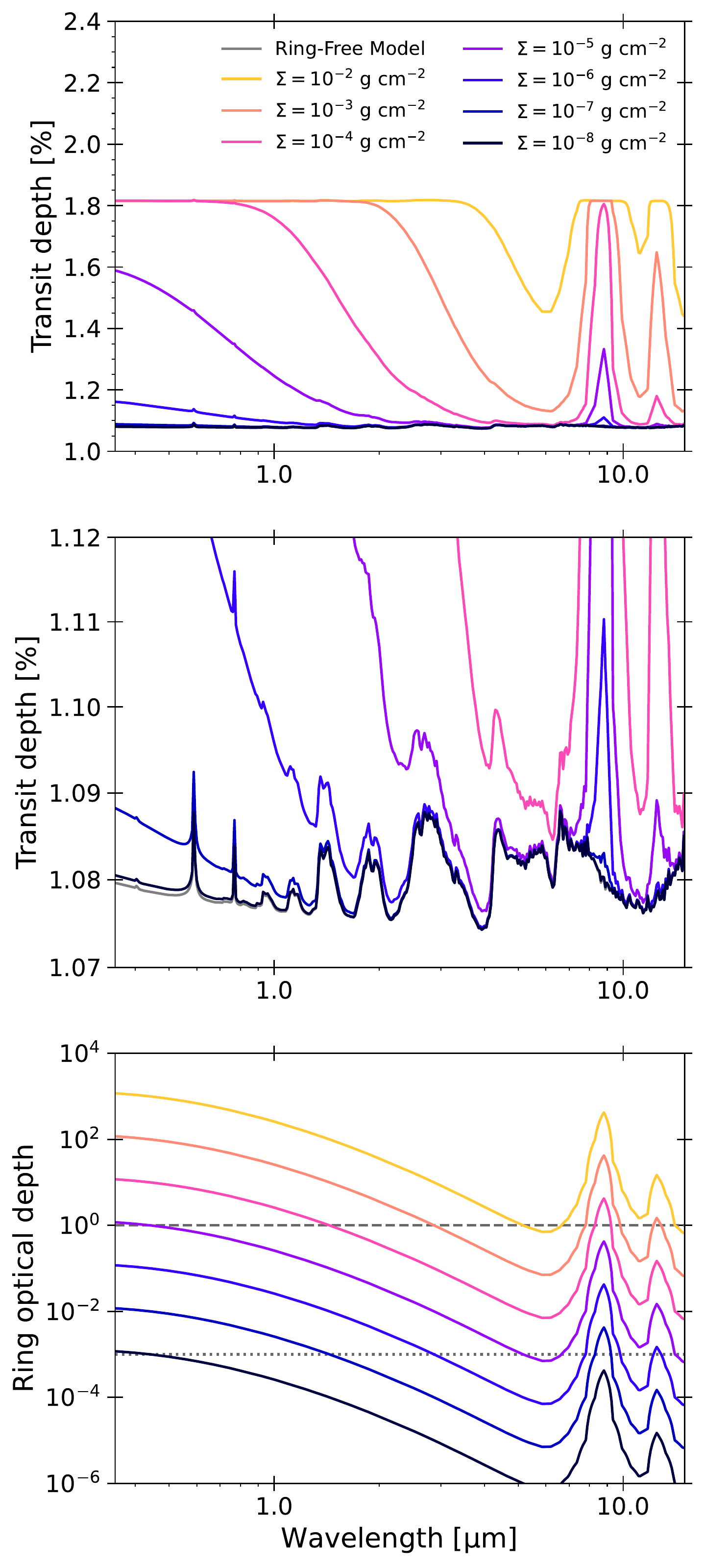}
\caption{
(Top) Transmission spectra of a Jupiter-mass planet with a CPDR. Different colored lines show the spectra for different ring surface densities. We assume optical properties of SiO$_2$ with a mean particle size and size distribution width of the CPDR as $a_{\rm r}=0.1~{\rm {\mu}m}$ and $\sigma=0.5$. The ring's inclination and width are fixed to $\phi=30~{\rm deg}$ and $W=0.5$. (Middle) The same as the top panel, but with a different vertical scale. (Bottom) The normal optical depth of the ring. The gray dashed and dotted lines denote the optical depth of $1$ and ${10}^{-3}$ to indicate the optically thin regime.
}
\label{fig:ring_spec}
\end{figure}

\begin{figure*}[t]
\centering
\includegraphics[clip, width=0.98\hsize]{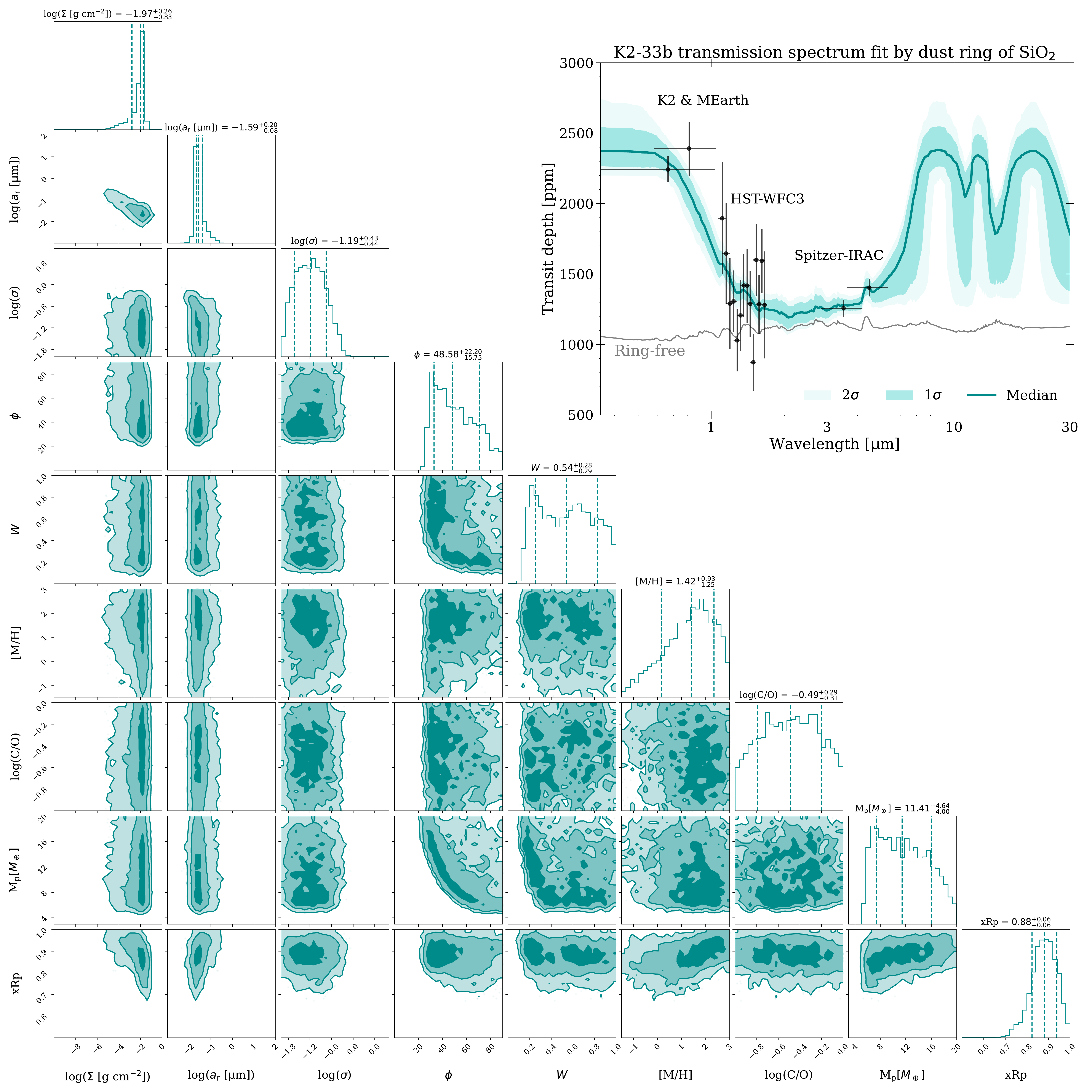}
\caption{Corner plot showing the results of atmosphere-ring retrieval on the transmission spectrum of K2-33b obtained by Thao et al. (2022). The colored regions in the 2D posteriors denote $1\sigma$, $2\sigma$, and $3\sigma$ intervals. The right top panel also shows the median transmission spectrum and its $1\sigma$ and $2\sigma$ uncertainties. The CPDR model succeeds in explaining a factor of $2$ variation in the transit depth from optical to near-infrared wavelength.  Prominent absorption features across the range from 8-30 $\mu$m are due to ring particle absorption.
}
\label{fig:ring_retrieval}
\end{figure*}
\begin{figure*}[t]
\centering
\includegraphics[clip, width=\hsize]{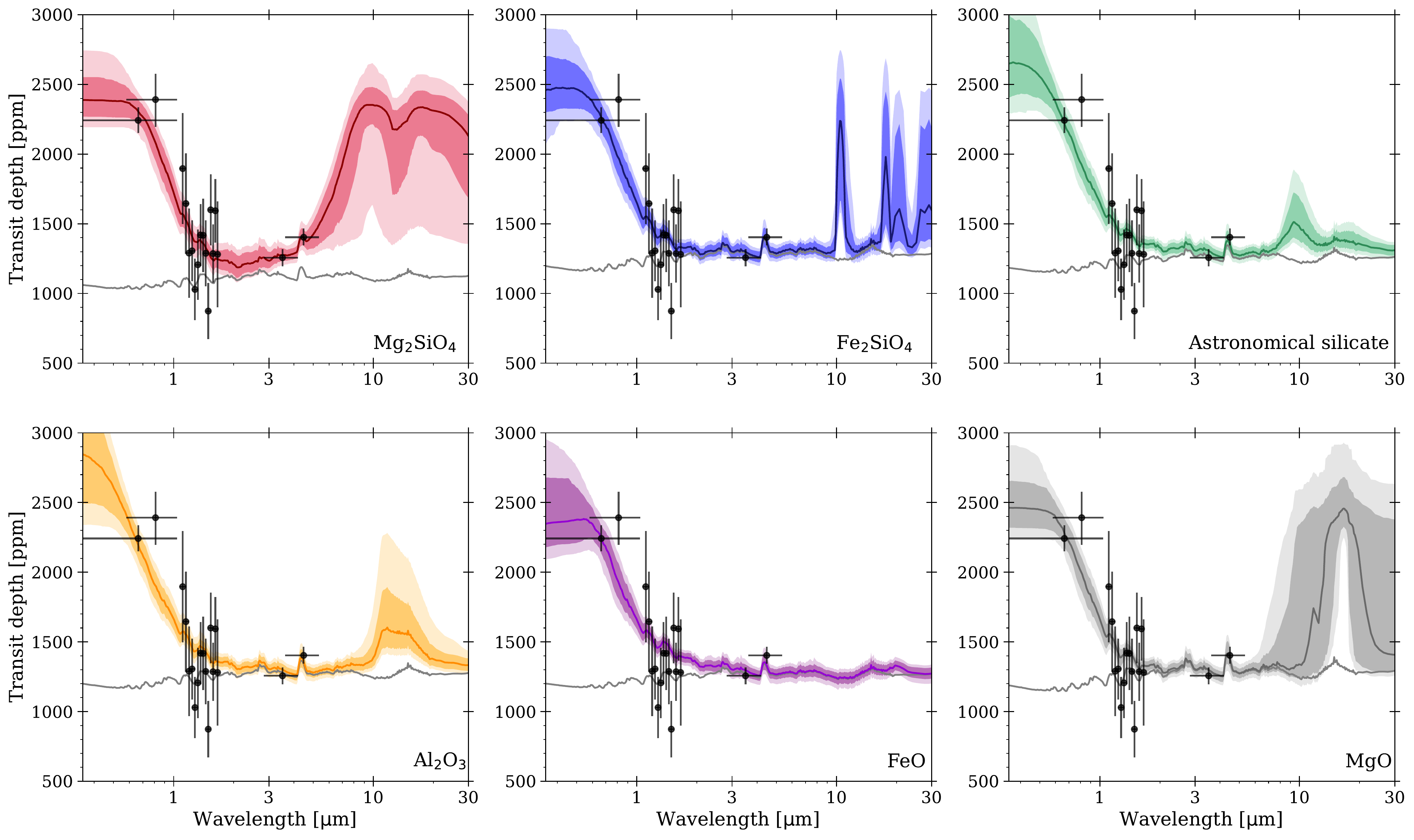}
\caption{Transmission spectra of K2-33b with various mineral compositions of the CPDR compared to a ring-free model (grey line). Importantly, all mineral compositions show prominent absorption features at $10$--$30~{\rm {\mu}m}$, except for FeO.}
\label{fig:spectrum_multi}
\end{figure*}

\subsection{Exoplanetary Transmission Spectra with Circumplanetary Dusty Ring}

We numerically demonstrate that an optically thin CPDR could produce an extreme spectral slope.
We first compute the synthetic ring-free transmission spectrum of a hypothetical K2-33b-like planet using a publicly available radiative transfer code, \texttt{CHIMERA} \citep{Line+13}.
We consider a hypothetical Jupiter-mass planet around a Sun-like star with an isothermal solar composition atmosphere of $T=1000~{\rm K}$ under thermochemical equilibrium.
Then we include the effect of the ring following a post-processing prescription described in \citet{Ohno&Fortney22}.
The outer edge (R$_{out}$) of the ring is fixed to the Roche radius with a particle density of $\rho_{\rm r}=3~{\rm g~{cm}^{-3}}$.
For the inner edge (R$_{in}$), we introduce the ring width parameter ($W$) as
\begin{equation}
    R_{\rm in} = (1-W)R_{\rm out}.
\end{equation}
We also vary the inclination of a ring $\phi=\pi/2 - i_{\rm r}$, where $i_{\rm r}$ is the inclination of the ring plane with respect to the sky plane \citep[see also][]{Akinsanmi+20}.
For example, the face-on and edge-on rings have angles of $\phi=90$ and $0~{\rm deg}$, respectively.
We computed the ring opacity using the publicly available Mie theory code \texttt{PyMieScatt} \citep{Sumlin+18}. 
We assume a log-normal particle size distribution given by
\begin{equation}
\frac{d\mathcal{N}}{da}=\frac{\mathcal{N}}{\sqrt{2\pi}\sigma a}\exp{\left[ -\frac{\ln^2{(a/a_{\rm r})}}{2\sigma^2} \right]},
\end{equation}
where $\mathcal{N}$ is the column number density of the ring particles, which is associated with the surface density as $\Sigma = \int (4\pi a^3\rho_{\rm r}/3)(d\mathcal{N}/da)da$.
In this study, we adopt the SiO$_2$ optical constants for a fiducial model and test other various optical constants in Section \ref{sec:spectrum_K2_33b}, as the composition of the exoplanetary ring is unknown.
We use refractive indices compiled by \citet{Kitzmann&Heng18} unless otherwise noted.

Figure \ref{fig:ring_spec} shows the spectrum with CPDR for various ring surface densities.
The corresponding ring optical depth is also shown in the bottom panel.
The ring produces a flat spectrum with a transit depth much larger than the ring-free spectrum when the ring is optically thick, in agreement with previous studies \citep{Ohno&Fortney22,Alam+22}.
Once the ring becomes optically thin, the transit depth rapidly drops to the depth of the ring-free planet.
This sudden drop appears as an extreme spectral slope in the transmission spectrum, as seen in the middle panel of Figure \ref{fig:ring_spec}.
The ring effect eventually becomes negligible at $\Sigma \la {10}^{-8}~{\rm g~{cm}^{-2}}$, equivalent to $\tau_{\rm ring}\la {10}^{-3}$ (see bottom panel).
These results are consistent with the analytic prediction presented in the previous section.

The CPDR produces significant absorption features around $9$ and $12.5~{\rm {\mu}m}$.
The amplitude of the feature depends on the ring surface density and thus optical depth.
In some cases, the feature shows a top-hat shape rather than a sharp absorption peak.
This stems from the fact that the ring becomes optically thick around the absorption feature, which results in the transit depth being limited by the physical size of the ring.
This top-hat shape of the absorption feature could be a unique signature of CPDR.

\section{Application to Transmission Spectrum of K2-33b}\label{sec:spectrum_K2_33b}
We investigate whether the CPDR model can explain the transmission spectrum of K2-33b.
Since many parameters have been uncertain, we conduct a joint atmosphere-ring retrieval on the transmission spectrum of K2-33b presented by Thao et al. (2022) that constrains both atmospheric and ring properties simultaneously in a Bayesian framework.
We combine the chemically-consistent atmospheric retrieval of \texttt{CHIMERA} with the post-processing ring prescription of \citet{Ohno&Fortney22} to perform this new atmosphere-ring retrieval.
The parameters retrieved are the ring surface density $\Sigma$, mean particle size $a_{\rm r}$, size distribution width $\sigma$, inclination $\phi$, width $W$, atmospheric metallicity [M/H], C/O ratio, planetary mass $M_{\rm p}$, and 10 bar radius $xR_{\rm p}R_{\rm p0}$, where $R_{\rm p0}=0.03545R_{\rm *}$ is the radius of K2-33b at Spitzer bands (Thao et al. 2022). 
Note that the planetary mass is also a free parameter since the current observations could only put an upper limit on the mass of $3.6M_{\rm jup}$ due to the large spot-induced radial velocity (RV) jitter \citep{Mann+16,David+16}.
We adopt the stellar properties of K2-33 described by Thao et al. (2022): $M_{\rm *}$= 0.571 $M_{\rm \odot}$, $R_{\rm *}$ = 1.017 $R_{\rm \odot}$, and $T_{\rm eff}$ = 3540 K.
For simplicity, we fix the atmospheric pressure-temperature profile to that calculated by an analytical model of \citet{Guillot10} with atmospheric infrared opacity of $\kappa_{\rm IR}=0.01~{\rm {cm}^2~g^{-1}}$, visible-to-infrared opacity ratio of $0.2$, planetary equilibrium temperature of $T_{\rm eq}=768~{\rm K}$ and intrinsic temperature of $T_{\rm int}=100~{\rm K}$.
We apply the Nested Sampling method with \texttt{pymultinest} \citep{Buchner+14}, which is a Python implementation of \texttt{MULTINEST} \citep{Feroz+09}, to estimate the posterior distribution of each parameter.
We adopt uniform prior distributions and 500 livepoints.

We summarize the retrieval result for the SiO$_2$ ring in Figure \ref{fig:ring_retrieval}.
The CPDR model could reasonably explain the overall transmission spectrum of K2-33b ($\chi^2=16.85$, the number of data points is $18$), including the sharp drop in transit depth from optical to near-infrared wavelength. 
The model retrieved a sub-micron particle size so that the CPDR could explain the extreme optical slope.
The spectrum features an equal contribution by the atmosphere and a ring at near-infrared wavelengths. 
The \textit{Spitzer} data points at 3.6 and 4.5 $\rm{\mu}m$ lead to a preference for a higher metallicity ($\sim30\times$ solar) and a sub-solar C/O so that the CO$_2$ and/or CO features dominate over the CH$_4$ feature, although the uncertainty is large. 
Future observations with wider wavelength coverage and better precision, along with the observational constraint on planetary mass, would provide better insight into the atmospheric properties of K2-33b.

It is worth noting that the CPDR model prefers a relatively high planetary mass, say $\ga 7M_{\rm \oplus}$ (see the posterior in Figure \ref{fig:ring_retrieval}),  compared to the aerosol scenario.
This is because a more massive planet can have a larger Roche radius.
This trend is opposed to the aerosol scenario, which instead prefers a lower planetary mass ($<5M_{\oplus}$; Thao et al. 2022), as the aerosol scenario requires a large atmospheric scale height to cause a large variation in the transit depth.
The preference for a high planetary mass is compatible with the requirement for atmospheric stability suggested by \citet{Kubyshkina+18}.
All CPDR models also predict silicate features with amplitudes of up to $\sim1000~{\rm ppm}$.
Therefore, the presence or absence of the feature directly tests whether a CPDR causes the extreme optical slope of K2-33b.
Such observations, which would be accessible by JWST-MIRI, would provide further information on the atmospheric and surrounding environment of the young exoplanet K2-33b.

While we have assumed the optical constants for SiO$_2$ for the CPDR, the actual composition of the CPDR remains unknown.
The CPDRs in the Solar System are possibly formed via impact ejecta from their satellites \citep[e.g.,][]{Burns+99,Burns+01}.
Thus, if the CPDR does exist, we anticipate that the composition of CPDR is similar to the crustal composition of parent bodies, e.g., exomoons and/or remaining satellitesimals.
If true, the assumption of SiO$_2$ may be reasonable, as the crusts of rocky objects in the inner Solar System primarily consist of SiO$_2$ ($\sim46$--$66$ wt\%, \citealt{McLennan22} and references therein).
We investigate the impact of different compositions of the CPDR by repeating the retrieval with various optical constants: representative olivines, which includes Mg$_2$SiO$_4$ and Fe$_2$SiO$_4$; moderately abundant crustal minerals, which includes Al$_2$O$_3$, FeO, and MgO; and astronomical silicate \citep{Draine03}.
As shown in Figure \ref{fig:spectrum_multi}, we find that all the mineral compositions tested here could explain the extreme optical slope of K2-33b while each mineral produces a distinct spectral feature around $\sim10~{\rm {\mu}m}$, except for FeO.
We note that a pure FeO composition is likely unrealistic, as the mass fraction of FeO is up to $\sim20~{\rm wt\%}$ in Solar System crustal rocks \citep{McLennan22} and predicted to be $\sim50~{\rm wt\%}$ (the reminders are mostly SiO$_2$ and MgO) even in a hypothetical coreless planet where the iron core did not form  \citep[][]{Elkins-Tanton&Seager08}.
Thus, similar to the spectroscopy of dust tails of disintegrating rocky exoplanets \citep{Bodman+18,Okuya+20}, future observations of absorption features at the mid-infrared wavelength would also provide a unique opportunity to explore the origins of CPDRs and their possible parent bodies.

\section{Summary and Discussion}\label{sec:summary}
We have shown that the optically thin CPDR could explain the extreme spectral slope of the $11~{\rm Myr}$ young exoplanet K2-33b.
The advantage of the CPDR scenario is that it prefers a high planetary mass (say, $\ga7M_{\rm \oplus}$), which is compatible with the requirement of atmospheric stability \citep{Kubyshkina+18}.
Our model predicts that the CPDR exhibits a prominent silicate feature at $\sim10~{\rm {\mu}m}$, which would be testable by JWST-MIRI.
It is important to disentangle which haze or CPDR scenario is true from future observations.
If the haze with a lower planetary mass scenario is true, it may challenge the current understanding of atmospheric escape, which is essential to understand overall planetary evolution.
Exoplanetary rings have not been conclusively detected yet, though several studies have suggested a few candidates \citep[for review, see][]{Heller18}, such as a putative ringed planet J1407b \citep[e.g.,][]{Mamajek+12,VanWerkhoven+14,Kenworthy&Mamajek15}, and a cool extremely low-density giant planet HIP 41378f \citep{Santerne+19,Akinsanmi+20,Alam+22,Belkovski+22}.
Thus, the detection of the absorption feature of the CPDR, if it is, would make K2-33b a very promising candidate for the first exoplanetary ring system.
Since the CPDR is possibly maintained by impact ejecta from the planet's moon(s), if the CPDR scenario is true, the absorption feature of CPDR may provide a unique opportunity to study the composition of exomoons, providing insights into their formation and evolution processes.

A more general implication of this study is that CPDR might provide an explanation for some other peculiar transmission spectra that cannot be explained by atmospheric processes.
For example, a low-density super-Neptune HATS-8b shows an extreme spectral slope that is steeper than the nominal Rayleigh slope by a factor of $\sim27$ \citep{May+20}.
Since this extreme slope could not be fully explained by stellar spots and aerosols \citep{May+20}, a CPDR may provide an explanation.
Although it is unclear to what degree CPDRs are ubiquitous in the exoplanetary system, it is worth considering the possibility of these and other peculiar transmission spectra.

Since we have adopted a highly simplified CPDR model, further studies will be warranted on the formation processes of CPDRs around exoplanetary environments.
Even a moderately optically thin (e.g., $\tau_{\rm ring}\sim {10}^{-2}$) CPDR can noticeably affect the transmission spectrum. 
Thus, it would be important to investigate what the typical CPDR expected for exoplanets is from a theoretical standpoint.
Meanwhile, the ring may leave other observable signatures \citep[see e.g.,][]{Zuluaga+15,Santos+15,deMooij+17,Aizawa+17}, such as forward scattering in transit light curve \citep{Barnes&Fortney04} and radial velocity anomaly \citep{Ohta+09}.
Future investigations by theoretical and observational studies would help to understand the observational consequences of CPDRs and their implications on their formation and evolution processes.

\acknowledgments 
We thank Callie Hood for the helpful comments on the use of \texttt{CHIMERA}. The authors also thank Dave Charbonneau, Adam Kraus, Gregory Sloan, and Andrew Vanderburg for suggesting that the chromatic transit depths seen for K2-33\,b might be due to a planetary ring. K.O. was supported by a JSPS Overseas Research Fellowship.  J.J.F. is supported by an award from the Simons Foundation. A.W.M is supported by a grant from the NSF CAREER program (AST-2143763). 
P.C.T was supported by NSF Graduate Research Fellowship (DGE-1650116), NC Space Grant Graduate Research Fellowship, the Zonta International Amelia Earhart Fellowship, and the Jack Kent Cooke Foundation Graduate Scholarship.

\bibliography{reference}

\end{document}